\documentclass[showpacs,aps,graphicx,twocolumn]{revtex4}
\usepackage{graphicx}

\begin{document}
\title{Detection of the nonlocal atomic entanglement assisted with single photons}
\author{ Yu-Bo Sheng,$^{1,2}$\footnote{Email address:
shengyb@njupt.edu.cn} and Lan Zhou$^{2,3}$ }
\address{$^1$ Institute of Signal Processing  Transmission, Nanjing
University of Posts and Telecommunications, Nanjing, 210003, China\\
 $^2$ Key Lab of Broadband Wireless Communication and Sensor Network
 Technology, Nanjing University of Posts and Telecommunications, Ministry of
 Education, Nanjing, 210003, China\\
 $^3$College of Mathematics \& Physics, Nanjing University of Posts and Telecommunications, Nanjing,
210003, China}
\begin{abstract}
We present an efficient way for measuring the entanglement of the atoms.
Through the auxiliary single photons input-output process in cavity quantum electrodynamics (QED), the concurrence of the
atomic entanglement can be obtained according to the success probability of picking up the singlet states of the atoms. This protocol has three advantages: First,  we do not require the sophisticated controlled-not (CNOT) gates. Second, the distributed atoms are not required to intact with each other. Third,  the atomic entanglement can be distributed nonlocally, which provides its important applications in distributed quantum computation.
\end{abstract} 
 \maketitle 

Entanglement plays an important role in current quantum information processing (QIP)\cite{rmp}. Quantum teleportation\cite{teleportation}, quantum key distribution \cite{Ekert91},
quantum dense coding \cite{densecoding}, quantum secure direct
communication \cite{QSDC1,QSDC2} and other quantum information
protocols \cite{QSS1,QSS2,QSS3}, all rely on the entanglement between distant parties.
However, entanglement is difficult to characterize experimentally. Bell inequalities \cite{Bell},
entanglement witnesses \cite{witness} cannot provide satisfactory results of the entanglement because they disclose the entanglement of some states while fail for other states. Another indirect method for measuring the entanglement is the quantum state tomographic reconstruction \cite{tomographic1,tomographic2}. By reading out the 15 parameters, they can reconstruct the density matrix of a two-qubit state, but it is quite complicated.

 In the early work of Bennett \emph{et al.}, they proposed the way for measuring the quantify entanglement named entanglement of formation \cite{concurrence1}. In their protocol,
 an arbitrary two-qubit state with the density matrix $\rho$ can be described in terms of exactly calculable quantity, i. e. the concurrence (C). The concurrence of the two-qubit
 state can be defined as \cite{concurrence2,concurrence3}
 \begin{eqnarray}
 C(\rho)=max\{0,\lambda_{1},-\lambda_{2},-\lambda_{3},-\lambda_{4}\},
 \end{eqnarray}
with the $\lambda_{i}(i=1,2,3,4)$. $\lambda_{i}$ are the non-negative eigenvalues of the Hermitian matrix
$R=\sqrt{\sqrt{\rho}\widetilde{\rho}\sqrt{\rho}}$  in decreasing order. Here $\widetilde{\rho}=(\sigma_{y}\otimes\sigma_{y})\rho^{\ast}(\sigma_{y}\otimes\sigma_{y})$, and
$\sigma_{y}$ is the Pauli operator, and  $\rho^{\ast}$ is the complex conjugate of $\rho$.
If we consider an arbitrary two-qubit pure state of the form $|\phi\rangle=\alpha|0\rangle|0\rangle+\beta|0\rangle|1\rangle+\gamma|1\rangle|0\rangle+\delta|1\rangle|1\rangle$, with $|\alpha^{2}|+|\beta|^{2}+|\gamma|^{2}+|\delta|^{2}=1$, the concurrence can be described as
$C(|\phi\rangle)=|\langle\phi|\sigma_{y}\otimes\sigma_{y}|\phi^{\ast}\rangle=2|\alpha\delta-\beta\gamma|$.

In 2006, Walborn \emph{et al.}   reported their experiment about the determination of entanglement with a single
measurement \cite{concurrence4}. In their experiment, they demonstrated the measuring of the concurrence for a two polarized state of the form $\alpha|01\rangle+\beta|10\rangle$
and the concurrence can be described as $C=2|\alpha|\sqrt{1-|\alpha|^{2}}$. They require the hyperentanglement to complete the task.
Recently, the group of Cao also discussed the measurement of the concurrence for two-photon polarization entangled pure state with the help of cross-Kerr nonlinearity \cite{concurrence5,concurrence6}.

On the other hand, the cavity quantum electrodynamics (QED) has become an important platform to realize the QIP \cite{rmpqed}. Over past
decades, both excellent theatrical and experiments have focused on QED for QIP, such as the generation the single photons \cite{singlephoton}, the performance of the logic gate \cite{logicgate}, and so on \cite{trapped}. In the early work of QED, they should require the high-quality cavity and the strong coupling between the solid qubit and the cavity. Recently, the QIPs based on the low-Q cavity have been widely discussed. For example, An \emph{et al.} showed that with the help of Faraday rotation, the different polarized photons can obtain the different phase shift after they interact with the atoms which are trapped in a low-Q cavity \cite{fengmang1}. Subsequently,  a lot of excellent works for atoms were discussed, including the quantum teleportation \cite{fengmang3}, controlled teleportation \cite{cteleportationqip}, swapping \cite{swapping}, entanglement concentration \cite{atomconcentration}, and so on \cite{entanglementgeneration,logicgate1}.

Interestingly, inspired by the previous works of atoms and the measurement of the concurrence, we find that the Faraday rotation can also be used to perform the measurement of the atomic entanglement. Actually, in 2007, Romero \emph{et al.}  first discussed the direct measurement for concurrence of atomic two-qubit pure state \cite{atomcurrence}. In their protocol, they illustrate the protocol for Rydberg atoms crossing three-dimensional microwave cavities and confined ions in a linear paul trap. They require the controlled-not (CNOT) gate or the controlled-phase gate between two atoms to complete the task. In 2008, Lee \emph{et al.} presented the concurrence of a two-qubit cavity system with the help of flying atoms \cite{cavityconcurrence}. The measurement for concurrence based on trapped ions were also proposed \cite{yangrc}.

In this paper, we will propose the detection of the remote  atomic entanglement with the help of single photons. This protocol is quite different from  others. First, it does not require the sophisticated controlled-not (CNOT) gates and the high-Q cavities, which will greatly release the experimental complexity. Second, the distributed atoms are not required to intact with each other. Third,  the atomic entanglement can be distributed nonlocally, which provides its important applications in distributed quantum computation.

This paper is organized as follows: In Sec. 2, we first briefly describe the key element of this protocol, i. e. the basic principle of the
Faraday rotation. In Sec. 3, we will explain our protocol with a simple example. In Sec. 4, we will make a discussion about the practical imperfection in experiment. In Sec. 5, we will make a conclusion.

\section{Basic theoretical model of the photonic Faraday rotation}
In this section, we will first explain the basic principle of the photonic Faraday rotation.

\begin{figure}[!h]
\begin{center}
\includegraphics[width=4cm,angle=0]{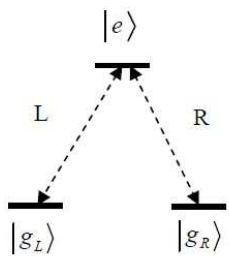}
\caption{The relevant atomic structure subjected to a low-Q cavity fielded. It has an excited state $|e\rangle$ and two degenerate ground states $|g_{L}\rangle$ and $|g_{R}\rangle$. It is also shown in Refs.\cite{cteleportationqip,atomconcentration}.}
\end{center}
\end{figure}
From Fig. 1, a three-level atom is trapped in the cavity. It has an excited state $|e\rangle$ and two degenerate ground states $|g_{L}\rangle$ and $|g_{R}\rangle$. The transitions between the  $|e\rangle$ and $|g_{L}\rangle$, $|e\rangle$ and $|g_{R}\rangle$ are assisted with a left-circularly ($|L\rangle$) and right-circularly ($|R\rangle$) polarized photon, respectively. We can write the Hamiltonian of the system as \cite{cteleportationqip}
\begin{eqnarray}
H&=&\sum_{j=L,R}[\frac{\hbar\omega_{0}\sigma_{jz}}{2}+\hbar\omega_{c}a^{\dagger}_{j}a_{j}]\nonumber\\
&+&\hbar\lambda\sum_{j=L,R}(a_{j}^{\dagger}\sigma_{j-}+a_{j}\sigma_{j+})+H_{R},
\end{eqnarray}
with
\begin{eqnarray}
H_{R}&=&H_{R0}+i\hbar[\int_{-\infty}^{\infty}d\omega\sum_{j=L,R}\alpha(\omega)(b_{j}^{\dagger}(\omega)a_{j}+b_{j}(\omega)a^{\dagger}_{j})]\nonumber\\
&+&\int_{-\infty}^{\infty}d\omega\sum_{j=L,R}\overline{\alpha}(\omega)(c_{j}^{\dagger}\sigma_{j-}+c_{j}\sigma_{j+}).
\end{eqnarray}
Here $a^{\dag}$ and $a$ are the creation and annihilation operators of the cavity field and the frequency of the cavity
field is $\omega_{c}$. The $\sigma_{R-}$, $\sigma_{R+}$, $\sigma_{L-}$  and $\sigma_{L+}$
are the lowering and raising operators of the transition R(L), respectively. The $\omega_{0}$ is the atomic frequency and $H_{R0}$ means the Hamiltonian of the free reservoirs.

Briefly speaking, we consider a single photon pulse with frequency $\omega_{P}$ enters  the optical cavity.  Using the adiabatic approximation, we can obtain the input-output relation of the cavity  field
in the form of \cite{entanglementgeneration,cteleportationqip,swapping}
\begin{eqnarray}
r(\omega_{p})=\frac{[i(\omega_{c}-\omega_{p})-\frac{\kappa}{2}][i(\omega_{0}-\omega_{p})+\frac{\gamma}{2}]
+\lambda^{2}}{[i(\omega_{c}-\omega_{p})+\frac{\kappa}{2}][i(\omega_{0}-\omega_{p})+\frac{\gamma}{2}]+\lambda^{2}}.\label{reflect1}
\end{eqnarray}
From Eq. (\ref{reflect1}), if $\lambda=0$, above equation for an empty cavity can be written as
\begin{eqnarray}
r_{0}(\omega_{p})=\frac{i(\omega_{c}-\omega_{p})-\frac{\kappa}{2}}{i(\omega_{c}-\omega_{p})+\frac{\kappa}{2}}.
\end{eqnarray}
Here $\omega_{p}$ is the frequency  of the photon. $\kappa$ and $\gamma$ are the cavity damping rate and atomic decay rate. The $\lambda$
is the atom-cavity coupling strength.
The total physical picture of the process can be described as follows: if a left circular polarization photon enters the cavity, and couples with
the atom in the state $|g_{L}\rangle$, the photon will be reflected to the output mode but induce a phase shift, which can be described as
$|L\rangle|g_{L}\rangle\rightarrow r(\omega_{p})|L\rangle|g_{L}\rangle\approx e^{i\theta}|L\rangle|g_{L}\rangle$. Here $e^{i\theta}=r(\omega_{p})$. Interestingly, if the right-circular photon enters the cavity and couples with  the atom in the state $|g_{L}\rangle$, it only senses the empty cavity. After the photon reflecting from the cavity, the
whole process can be written as $|R\rangle|g_{L}\rangle\rightarrow r_{0}(\omega_{p})|R\rangle|g_{L}\rangle\approx e^{i \theta_{0}}|R\rangle|g_{L}\rangle$. Here $e^{i \theta_{0}}=r_{0}(\omega_{p})$.
Therefore, if a single photon state $\frac{1}{\sqrt{2}}(|L\rangle+|R\rangle)$ enters the cavity and couples with the $|g_{L}\rangle$ atom, the whole system
can be described as
\begin{eqnarray}
\frac{1}{\sqrt{2}}(|L\rangle+|R\rangle)|g_{L}\rangle \rightarrow \frac{1}{\sqrt{2}}( e^{i\theta}|L\rangle+e^{i \theta_{0}}|R\rangle)|g_{L}\rangle.
\end{eqnarray}
On the other hand, if  $\frac{1}{\sqrt{2}}(|L\rangle+|R\rangle)$ enters the cavity and couples with the $|g_{R}\rangle$ atom, the whole system
can be described as
\begin{eqnarray}
\frac{1}{\sqrt{2}}(|L\rangle+|R\rangle)|g_{R}\rangle \rightarrow \frac{1}{\sqrt{2}}( e^{i\theta_{0}}|L\rangle+e^{i \theta}|R\rangle)|g_{R}\rangle.
\end{eqnarray}
The $\Delta\Theta_{F}=|\theta-\theta_{0}|$  is called the Faraday rotation.

If we consider the cases that $\omega_{0}=\omega_{c}$, $\omega_{p}=\omega_{c}-\frac{\kappa}{2}$, and $\lambda=\frac{\kappa}{2}$, we can
obtain $\theta=\pi$ and $\theta_{0}=\frac{\pi}{2}$. We can rewrite the relationship between the photon and the atom as
\begin{eqnarray}
|L\rangle|g_{L}\rangle\rightarrow-|L\rangle|g_{L}\rangle, |R\rangle|g_{L}\rangle\rightarrow i|R\rangle|g_{L}\rangle,\nonumber\\
|L\rangle|g_{R}\rangle\rightarrow i|L\rangle|g_{R}\rangle, |R\rangle|g_{R}\rangle\rightarrow -|R\rangle|g_{R}\rangle.\label{relation}
\end{eqnarray}
\begin{figure}[!h]
\begin{center}
\includegraphics[width=7cm,angle=0]{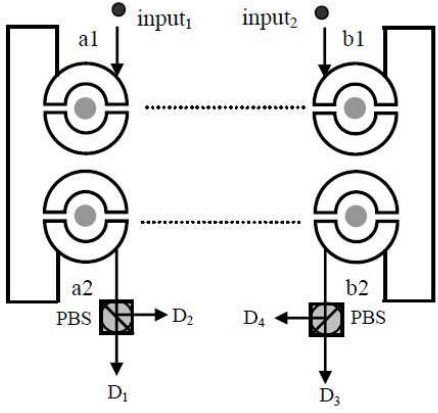}
\caption{The schematic drawing of the principle of our protocol. Two input polarized photons are in the same state $\frac{1}{\sqrt{2}}(|L\rangle+|R\rangle)$. The polarization beam splitter can transmit the photon in the $|+\rangle$ polarization and reflect the photon in the $|-\rangle$ polarization.
Here $|\pm\rangle=\frac{1}{\sqrt{2}}(|L\rangle\pm|R\rangle)$. D1, D2, D3 and D4  are four single-photon detectors.}
\end{center}
\end{figure}
\section{Measuring the concurrence of the atomic entanglement with single photons}
Now we start to explain our protocol. As shown in Fig. 1, two pairs of atomic entanglement are shared by Alice and Bob, respectively.
Suppose that they are in the same arbitrary two-qubit states of the form
 \begin{eqnarray}
 |\Phi\rangle=\alpha|0\rangle_{a}|0\rangle_{b}+\beta|0\rangle_{a}|1\rangle_{b}+\gamma|1\rangle_{a}|0\rangle_{b}+\delta|1\rangle_{a}|1\rangle_{b}.
  \end{eqnarray}
Here $|g_{L}\rangle\equiv|0\rangle$ and $|g_{R}\rangle\equiv|1\rangle$. First, they prepare two single photons with the same polarization $|\Phi_{1}\rangle=|\Phi_{2}\rangle=\frac{1}{\sqrt{2}}(|L\rangle+|R\rangle)$. They let the two photons in the intput1
and input2 pass through the cavities, respectively.

The four atoms can be written as
 \begin{eqnarray}
 &&|\Phi\rangle_{a1b1}\otimes|\Phi\rangle_{a2b2}\nonumber\\
&=&(\alpha|0\rangle_{a1}|0\rangle_{b1}+\beta|0\rangle_{a1}|1\rangle_{b1}+\gamma|1\rangle_{a1}|0\rangle_{b1}+\delta|1\rangle_{a1}|1\rangle_{b1})\nonumber\\
 &\otimes&(\alpha|0\rangle_{a2}|0\rangle_{b2}+\beta|0\rangle_{a2}|1\rangle_{b2}+\gamma|1\rangle_{a2}|0\rangle_{b2}+\delta|1\rangle_{a2}|1\rangle_{b2})\nonumber\\
  &=&\alpha^{2}|0\rangle_{a1}|0\rangle_{a2}|0\rangle_{b1}|0\rangle_{b2}+\alpha\beta|0\rangle_{a1}|0\rangle_{a2}|0\rangle_{b1}|1\rangle_{b2}\nonumber\\
  &+&\alpha\gamma|0\rangle_{a1}|1\rangle_{a2}|0\rangle_{b1}|0\rangle_{b2}+\alpha\delta|0\rangle_{a1}|1\rangle_{a2}|0\rangle_{b1}|1\rangle_{b2}\nonumber\\
  &+&\alpha\beta|0\rangle_{a1}|0\rangle_{a2}|1\rangle_{b1}|0\rangle_{b2}+\beta^{2}|0\rangle_{a1}|0\rangle_{a2}|1\rangle_{b1}|1\rangle_{b2}\nonumber\\
  &+&\beta\gamma|0\rangle_{a1}|1\rangle_{a2}|1\rangle_{b1}|0\rangle_{b2}+\beta\delta|0\rangle_{a1}|1\rangle_{a2}|1\rangle_{b1}|1\rangle_{b2}\nonumber\\
  &+&\alpha\gamma|1\rangle_{a1}|0\rangle_{a2}|0\rangle_{b1}|0\rangle_{b2}+\beta\gamma|1\rangle_{a1}|0\rangle_{a2}|0\rangle_{b1}|1\rangle_{b2}\nonumber\\
  &+&\gamma^{2}|1\rangle_{a1}|1\rangle_{a2}|0\rangle_{b1}|0\rangle_{b2}+\gamma\delta|1\rangle_{a1}|1\rangle_{a2}|0\rangle_{b1}|1\rangle_{b2}\nonumber\\
  &+&\alpha\delta|1\rangle_{a1}|0\rangle_{a2}|1\rangle_{b1}|0\rangle_{b2}+\beta\delta|1\rangle_{a1}|0\rangle_{a2}|1\rangle_{b1}|1\rangle_{b2}\nonumber\\
  &+&\gamma\delta|1\rangle_{a1}|1\rangle_{a2}|1\rangle_{b1}|0\rangle_{b2}+\delta^{2}|1\rangle_{a1}|1\rangle_{a2}|1\rangle_{b1}|1\rangle_{b2}.\label{odd1}
  \end{eqnarray}
 Therefore, after the two photons passing through the four cavities, if both photons do not change and transmit the polarization beam splitters (PBSs) and finally are detected
 by single photon detectors D$_{1}$ and  D$_{3}$, the atomic state will become
  \begin{eqnarray}
  |\Phi\rangle_{a1a2b1b2}
   &=&\frac{\alpha\delta}{\sqrt{2(|\alpha\delta|^{2}+|\beta\gamma|^{2})}}(|0\rangle_{a1}|1\rangle_{a2}|0\rangle_{b1}|1\rangle_{b2}\nonumber\\
   &+&|1\rangle_{a1}|0\rangle_{a2}|1\rangle_{b1}|0\rangle_{b2})\nonumber\\
  &+&\frac{\beta\gamma}{\sqrt{2(|\alpha\delta|^{2}+|\beta\gamma|^{2})}}(|0\rangle_{a1}|1\rangle_{a2}|1\rangle_{b1}|0\rangle_{b2}\nonumber\\
  &+&|1\rangle_{a1}|0\rangle_{a2}|0\rangle_{b1}|1\rangle_{b2}),\label{step1}
    \end{eqnarray}
  with the total success probability of $P_{1}=2|\alpha\delta|^{2}+2|\beta\gamma|^{2}$.
After they obtain the state $  |\Phi\rangle_{a1a2b1b2}$ in Eq. (\ref{step1}), they perform the Hadamard operations on the atoms
a1 and a2, which makes the $|\Phi\rangle_{a1a2b1b2}$ become
  \begin{eqnarray}
   |\Phi\rangle'_{a1a2b1b2}&=&\frac{\alpha\delta+\beta\gamma}{2\sqrt{2(|\alpha\delta|^{2}+|\beta\gamma|^{2})}}(|0\rangle_{a1}|0\rangle_{a2}-|1\rangle_{a1}|1\rangle_{a2})\nonumber\\
   &\otimes&(|0\rangle_{b1}|1\rangle_{b2}+|1\rangle_{b1}|0\rangle_{b2})\nonumber\\
  &+&\frac{\alpha\delta-\beta\gamma}{2\sqrt{2(|\alpha\delta|^{2}+|\beta\gamma|^{2})}}(|0\rangle_{a1}|1\rangle_{a2}-|1\rangle_{a1}|0\rangle_{a2})\nonumber\\
  &\otimes&(|0\rangle_{b1}|1\rangle_{b2}-|1\rangle_{b1}|0\rangle_{b2}).\label{step2}
    \end{eqnarray}
From Eq. (\ref{step2}), after performing the Hadamard operation, the atoms a1 and a2 are entangled. Finally, they still prepare a single photon in the input1 mode of the form $\frac{1}{\sqrt{2}}(|L\rangle+|R\rangle)$, and let it pass through the two cavities
a1 and a2. If the photon's polarization does not change, it will project the atoms a1 and a2 into the singlet state $|0\rangle_{a1}|1\rangle_{a2}-|1\rangle_{a1}|0\rangle_{a2}.$ Therefore, if the photon passes through the cavities and finally be detected by the single-photon detector D$_{1}$ in a second time, the state $ |\Phi\rangle'_{a1a2b1b2}$ will collapse to
\begin{eqnarray}
 |\Phi\rangle''_{a1a2b1b2}&=&\frac{1}{2}(|0\rangle_{a1}|1\rangle_{a2}-|1\rangle_{a1}|0\rangle_{a2})\nonumber\\
 &\otimes&(|0\rangle_{b1}|1\rangle_{b2}-|1\rangle_{b1}|0\rangle_{b2}).
    \end{eqnarray}
The success probability for obtaining the state $ |\Phi\rangle''_{a1a2b1b2}$ is $P_{2}=\frac{|\alpha\delta-\beta\gamma|^{2}}{2(|\alpha\delta|^{2}+|\beta\gamma|^{2})}$.

The total success probability for obtaining the state $ |\Phi\rangle''_{a1a2b1b2}$ is $P=P_{1}P_{2}=|\alpha\delta-\beta\gamma|^{2}$.
Therefore, we can get
\begin{eqnarray}
C(\Phi)=2|\alpha\delta-\beta\gamma|=2\sqrt{P}.\label{c}
    \end{eqnarray}

    \begin{figure}[!h]
\begin{center}
\includegraphics[width=8cm,angle=0]{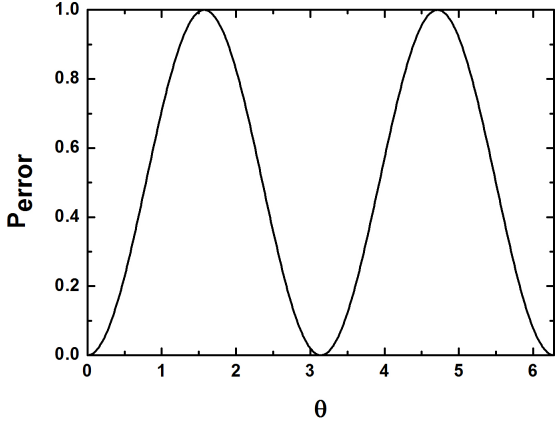}
\caption{The probability of error is altered with  $\theta$. We let $\theta\in (0,2\pi)$.}
\end{center}
\end{figure}
Interestingly, from Eq. (\ref{c}), we have established the relationship between the concurrence of the atomic entanglement and the success probability for
measuring the photons. In a practical experiment, we should repeat this protocol for many times to obtain the success probability $P$ by calculating the ratio
between the detected photon number and the initial total photon number.

\section{Discussion and conclusion}
So far, we have briefly described our protocol with a simple example. It is shown that we can complete the detection of the concurrence of  arbitrary
two-qubit atomic pure state by input-output processing of photons. In the pioneer work of Romero \emph{et al.},  they also discussed the measurement for concurrence of atomic two-qubit pure states. They require the atomic  CNOT gate to complete the task. However, the CNOT gate cannot be well performed in current
experimental conditions.  In this protocol, we do not require the CNOT gate, and we also do not require the atoms to interact with each other. In our protocol, $^{87}$Rb atom trapped in a fiber-based
Fabry-Perot cavity is a good candidate system \cite{atom}. In Ref.\cite{atom}, the two ground states $|g_{L}\rangle$ and $|g_{R}\rangle$ can be the states of $|F=2\rangle,m_{F}=\pm1$ of level $5S_{1/2}$.
The transition frequency between the excited state and the ground state is $\omega_{0}=2\pi c/\lambda$ at $\lambda=780nm$.  the cavity length $L=38.6\mu m$. The cavity decay rate $\kappa=2\pi\times53$ MHz, and the finesse $F=37000$. Certainly, in a practical experiment, we also should consider the experimental imperfection. For example, in this protocol, we require
$\omega_{0}=\omega_{c}$,  $\omega_{p}=\omega_{c}-\frac{\kappa}{2}$, and $\lambda=\frac{\kappa}{2}$. Actually, in practical experiment, we cannot ensure to control $\theta=\pi$ to let $\lambda=\frac{\kappa}{2}$. On the other hand, the photon loss and the imperfect detection will also affect the quality of the whole protocol. If $\theta\neq\pi$, we should rewrite the relationship in Eq. (\ref{relation}) as
\begin{eqnarray}
|L\rangle|g_{L}\rangle\rightarrow e^{i\theta}|L\rangle|g_{L}\rangle, |R\rangle|g_{L}\rangle\rightarrow i|R\rangle|g_{L}\rangle,\nonumber\\
|L\rangle|g_{R}\rangle\rightarrow i|L\rangle|g_{R}\rangle, |R\rangle|g_{R}\rangle\rightarrow e^{i\theta}|R\rangle|g_{R}\rangle.\label{relation1}
\end{eqnarray}
We can also obtain that
\begin{eqnarray}
\frac{1}{\sqrt{2}}(|L\rangle+|R\rangle)|0\rangle|0\rangle&\rightarrow&\frac{1}{\sqrt{2}}( e^{i2\theta}|L\rangle-|R\rangle)|0\rangle|0\rangle,\nonumber\\
\frac{1}{\sqrt{2}}(|L\rangle+|R\rangle)|1\rangle|1\rangle&\rightarrow&\frac{1}{\sqrt{2}}( -|L\rangle+e^{i2\theta}|R\rangle)|1\rangle|1\rangle,\nonumber\\
\frac{1}{\sqrt{2}}(|L\rangle+|R\rangle)|0\rangle|1\rangle&\rightarrow&\frac{1}{\sqrt{2}} ie^{i\theta}(|L\rangle+|R\rangle)|0\rangle|1\rangle,\nonumber\\
\frac{1}{\sqrt{2}}(|L\rangle+|R\rangle)|1\rangle|0\rangle&\rightarrow&\frac{1}{\sqrt{2}} ie^{i\theta}( |L\rangle+|R\rangle)|1\rangle|0\rangle.\label{relation3}
 \end{eqnarray}
From above description, it is shown that if the atomic state is $|0\rangle|1\rangle$ or  $|1\rangle|0\rangle$, the polarization state is not changed. That is to say, the imperfect operation does not affect the fidelity of the state. However, if the atomic state is $|0\rangle|0\rangle$ or  $|1\rangle|1\rangle$,
it will make the polarization of the photon change, which will has some probability to be $\frac{1}{\sqrt{2}}(|L\rangle+|R\rangle)$, and finally be detected by
D$_{1}$ or D$_{3}$. In this way, it will contribute a successful case with the probability of
\begin{eqnarray}
P_{error}&=&|\frac{1}{\sqrt{2}}(\langle L|+\langle R|)\frac{1}{\sqrt{2}}( e^{i2\theta}|L\rangle-|R\rangle)|^{2}\nonumber\\
&=&\frac{|e^{i2\theta}-1|^{2}}{4}.
 \end{eqnarray}
In Fig. 3, we calculate the $P_{error}$ altered with $\theta$.  From Eq. (\ref{odd1}), in an ideal case, the item $|0\rangle_{a1}|0\rangle_{a2}|0\rangle_{b1}|0\rangle_{b2}$ has no contribution to the successful case, while
it will has the probability of $|\alpha|^{4}P^{2}_{error}$ to make D$_{1}$ and D$_{3}$ both fire. The second item $|0\rangle_{a1}|0\rangle_{a2}|0\rangle_{b1}|1\rangle_{b2}$ also has the probability of $|\alpha\beta|^{2}P_{error}$ to make  D$_{1}$ and D$_{3}$ both fire.
Therefore, if $P_{error}\ll1$, we can estimate the total success probability as
\begin{eqnarray}
P'\approx|\alpha\delta-\beta\gamma|^{2}(1+P_{error})^{2}.
 \end{eqnarray}
 Here we omit the contribution of $P^{2}_{error}$ for $P^{2}_{error}\ll P_{error}$. Therefore, the concurrence will be increased because of the imperfection.
  On the other hand, we should also consider the photon loss. The photon loss is the main obstacle in  realistic experiment. The  minor misalignment, cavity mirror absorption and scattering, and even the tiny piece of dust, can induce the photon loss. The photon loss will decrease the total success probability. Suppose the efficiency of a single-photon detector is $\eta_{d}$, the efficiency of coupling and transmission of the photon from fiber to cavity is $T_{f}$, and the  transmission of each photon through the other optical components is $T_{o}$, we can rewrite the total success probability in a detection round as
 \begin{eqnarray}
 P_{T}=P'\times T^{3}_{f}\times T^{3}_{o}\times\eta^{3}_{d}.
 \end{eqnarray}
 In practical condition, the efficiency of the single-photon detector is $\eta_{d}=0.28$. Suppose that  $T_{f}=0.5$, and $T_{o}=0.95$. We can obtain that
 $ P_{T}\approx 2.4\times10^{-3}\times P'$. If the generation rate of the single-photon is $1\times 10^{-5}s^{-1}$, this protocol can be completed in 0.01s.

 In conclusion, we have described the detection of the nonlocal atomic entanglement assisted with single photons. Through the auxiliary single photons input-output process in cavity quantum electrodynamics (QED), the concurrence of the
atomic entanglement can be obtained according to the success probability of picking up the singlet states of the atoms. We also discussed the practical imperfection of this protocol. It is shown that the success probability will be increased if $\frac{\kappa}{2}\neq\lambda$. On the other hand, the photon loss will decrease the total success probability. This protocol has several advantages: First, we do not require the sophisticated CNOT gates. Second, the distributed atoms are not required to intact with each other. Third,  the atomic entanglement can be distributed nonlocally, which provides its important applications in distributed quantum computation. 

\section*{ACKNOWLEDGEMENTS}
This work is supported by the National Natural Science Foundation of
China under Grant Nos. 11104159 and 11347110.  and the Project
Funded by the Priority Academic Program Development of Jiangsu
Higher Education Institutions.

\end{document}